\documentclass[aps,twocolumn,showpacs]{revtex4-1}
\usepackage{graphicx}
\usepackage{textcomp}
\usepackage{psfrag}
\usepackage{epstopdf}
\usepackage{amssymb}
\usepackage{xcolor}
\usepackage{float}
\usepackage{lineno}
\usepackage{setspace}
\usepackage{lipsum}%
\usepackage{soul}

\begin{document}
\setstretch{1.3} 
\renewcommand\linenumberfont{\normalfont\scriptsize\sffamily\color{red}}
\title{A reductive analysis of a compartmental model for COVID-19: data assimilation and forecasting for the United Kingdom } 

\author{G. Ananthakrishna$^{1,*}$ and Jagadish Kumar$^{2,}$}
\email{Corresponding authors\\garani@iisc.ac.in\\
jagadish.physics@utkaluniversity.ac.in}
\affiliation{$^1$ Materials Research Centre, Indian Institute of Science, Bengaluru 560012, India\\
$^2$ Department of Physics, Utkal University, Bhubaneswar 751004, India 
}

%\textcolor{red}{}

%\linenumbers
\begin{abstract}
%\large 
We introduce a deterministic model that partitions the total population into the susceptible, infected, quarantined, and those traced after exposure, the recovered and the deceased. 
We hypothesize  'accessible population for transmission of the disease' to be a small fraction of the total population, for instance when interventions are in force. This hypothesis, together with the structure of the set of coupled nonlinear ordinary  differential equations for the populations, allows us to decouple the equations into just two equations. This further reduces to a logistic type of equation for the total infected population. The equation can be solved analytically and therefore allows for a clear interpretation of the growth and inhibiting factors in terms of the parameters in the full model. The validity of the 'accessible population' hypothesis and the efficacy of the reduced logistic model is demonstrated by the ease of fitting the United Kingdom data for the cumulative  infected and daily new infected cases. The model can also be used to forecast further progression of the disease. In an effort to find optimized parameter values compatible with the United Kingdom coronavirus data, we first determine the relative importance of the various transition rates participating in the original model. Using this we show that the original model equations provide a very good fit with the United Kingdom data for the cumulative number of infections and the daily new cases. The fact that the model calculated daily new cases  exhibits a turning point, suggests  the beginning of a slow-down in the spread of infections. However, since the rate of slowing down beyond the turning point is small, the cumulative number of infections is likely to saturate to about $3.52 \times 10^5$ around late July, provided the lock-down conditions continue to prevail. Noting that the fit obtained from the reduced logistic equation is comparable to that with the full model equations, the underlying causes for the limited forecasting ability of the reduced logistic equation are elucidated. The model and the procedure adopted here are expected to be useful in fitting the data for other countries and in forecasting the progression of the disease.

\end{abstract}
 
\maketitle

\section{Introduction} 
%\large
The highly contagious SARS-CoV-2 has infected more than six million people worldwide since its first detection in China on December 31 %{\color{red} \textbf{cite DOI:https://doi.org/10.1016/S1473-3099(20)30200-0}} 
\cite{Lescure20}. The novel coronavirus is the fourth wave in the class of coronaviruses. In less than two months, the virus has spread all over the world, posing serious threats to healthcare systems and economies. The alarming speed of transmission, the virulence of the disease, and the unprecedented high proportion of fatalities even in countries with high healthcare indices have raised questions about what kind of interventions are appropriate for a given setting. 
 The wide variability in infected numbers and fatalities in different countries and settings has also brought into sharp focus a debate about the underlying causes of the variability. In the absence of any treatment for the disease and non-availability of vaccines in the near future, policy makers have resorted to standard epidemiological interventions, such as social distancing, isolation, contact tracing, and quarantining, and more recently a complete lock-down. 

At a basic level, the purpose of all non-pharmacological interventions is to control disease transmission by limiting the proportion of the population exposed to the virus as much as possible. Furthermore, inherent in the process of implementation of these interventions are  delays at each stage. The delay time-scales are specific to the particular intervention.

The importance of mathematical models describing the spreading dynamics of infectious diseases has been recognized since early days \cite{Murray93}. In particular, the fact that timely  models that include realistic features have often been helpful in decision making on healthcare issues is well recognized \cite{Murray93,Kermack27,Egger17}. In the short period since the emergence of the coronavirus, there have been several mathematical models \cite{Tang20,Tang20a,Lin20,Zhao20,Chen20,Ferretti20,Hellewell20,Ferguson20,Dehning20,Salje20,Friston20,Muller20,Chinazzi20}, to name a few. Some of these models attempt to evaluate the contribution from different transmission routes, such as contact tracing and isolation \cite{Ferretti20,Hellewell20,Chavez03}, travel restrictions \cite{Muller20,Chinazzi20} social distancing \cite{Falco20,Mars20}, lock-down measures \cite{Muller20,Chinta20,Tobias20}, and a combination of these interventions \cite{Ferguson20,Eubank20,Tuite20,Chowdhury20,Sebastiani20}. These models broadly fall into three categories, deterministic, stochastic, and simulations. Several new mathematical techniques used in different disciplines have been employed to gain insights, which would not be possible with the traditional approaches in the field. These include the human mobility model \cite{Muller20}, differential evolution\cite{Falco20}, heuristic optimization technique \cite{Falco20}, stochastic agent-based discrete time simulation \cite{Chang20}, supply chain risk simulations \cite{Ivanov20}, etc. 
 
One class of epidemiological models attempts to describe the transmission dynamics by partitioning the population into smaller subsets based on the disease status such as the susceptible, exposed, infected, quarantined, recovered, etc \cite{Tang20,Tang20a,Lin20,Zhao20,Chen20}. Most models of this kind ignore age-dependent infection and fatality rates, and the heterogeneous spatial distribution of the population. In a sense, these models describe the evolution of the mean response of each type of population. Despite these limitations, these models have the ability to include several realistic features. 

 In the compartment type of models, the disease status of individuals changes with the development of the disease, i.e., transitions occur between two compartments either due to interaction of the infected with the susceptible or due to interventional actions. These models include delay time-scales inherent in the dynamics of transmission, for instance, the period spent in quarantine and the time required for tracing individuals exposed to the infected. These models have the ability to include several realistic features, such as the response of the population to interventional measures. However, generally, inclusion of more and more realistic features requires a larger number of partitions. Then, the number of differential equations increases and so does the number of parameters, making calibration of the parameters difficult \cite{Tang20,Tang20a,Lin20,Zhao20,Chen20,Dehning20}. 
 
Motivated by the complexity of such models, we have devised a compartment-based model having the susceptible, infected, quarantined, traced, recovered, and the deceased populations. The susceptible and the infected form the core populations in the sense that it is through these two populations that inward/outward transitions occur with other populations. However, even this simple model contains several parameters, leaving numerical solutions as the only option for further analysis. Considering the fact that the model equations form a coupled set of nonlinear differential equations, we adopt mathematical methods drawn from nonlinear dynamical  systems most appropriate for analyzing such equations \cite{Strogatz}. We hypothesize  'accessible population for infection' to be  a small fraction of the total population. The validity of this hypothesis can be appreciated by noting that the purpose of interventions is to minimize the exposure of the population to virus transmission, thereby limiting the spread of infection. We further assume that the order of magnitude of the accessible population is {\it similar to that of the infected population}. This assumption is made more quantitative. This, together with the structure of the model equations, allows us to decouple them into two equations. These two equations further reduce to a logistic type of equation for the total infected population with well defined parameters namely, the 'testing rate' and 'contact transmission rate' parameters \cite{Thyaga20,Kriston20}. The equation can be solved analytically, thereby allowing for a clear interpretation of the parameters controlling the growth and inhibiting factors. The validity of the 'accessible population' hypothesis and the efficacy of the reduced logistic model is demonstrated by the ease of fitting the cumulative number of infections and daily new cases for the United Kingdom (UK). The procedure further allows us to forecast the progression of the disease. Using this information and calibrating the relative importance of the various transition rates (equivalently the associated parameters), we optimize the parameter values specific to the UK. This procedure allows us to delineate the various time scales contributing to the various regions in the time development of the disease. Using this, we numerically solve the full model equations. The calculated total infected population fits very well with the available data for the UK \cite{WHO}. (UK does not publish data on the recovered and the active populations.) The model exhibits a turning point in the active infected population around May 15. However, since the rate of slowing down beyond the turning point is poor, the projected end time of the epidemic would be around late July with the predicted saturation level of the total number of infections $\sim 3.52 \times10^5$ assuming lock-down conditions continue.

\section{The model}
\label{F-model}

The total population $N$ is partitioned into the susceptible $S$, active infected $I$, quarantined $Q$, those traced $T$ after being exposed to the infected, recovered $R$, and the deceased $D$. The respective populations are denoted by $N_s, N_i,N_q,N_{tr},N_r$ and $N_d$.

Testing is one of the standard protocols used for identifying the infected. If $\alpha_s$ is the rate of testing per day per million and $p_s$ is the probability of testing positive, then $\alpha_s p_s N_s$ is the transition rate from $S$ to $I$. Infected individuals coming into contact with the susceptible class can transmit the virus. If $\beta_i$ is the transmission rate per contact, $p_i$ is the probability of transmission of the disease and $F(d_i)$ is a distance-dependent interaction, then, $ p_i \beta_i F(d_i) N_i N_s$ is the transition rate from $S$ to $I$. Considering the fact that one of the primary routes of transmission is through airborne aerosols generated by the infected, larger separation is known to reduce the risk of transmission \cite{Falco20,Mars20,Liu20}. 
 This distance-dependence of $F(d_i)$ is generally expressed as $F(d_i)$ $\propto$ $1/d^2$ or $1/d_i^3$. However, in the present context where we will be dealing with a lock-down situation for most part of the progression of the disease, we set $F(d_i) =1$.

 During testing, some individuals would always exhibit mild or ambiguous symptoms. These are identified as pre-symptomatic. If the probability of finding the pre-symptomatic is $p_q$, then, $\alpha_s p_q N_s$ transition out of $S$ to $Q$. Subsequently, when tested again, say after a quarantine duration \cite{Tang20,Becker20,Linton20}, some of them may either test positive with a probability $q_1$ or negative with a probability $(1-q_1)$. If positive, the transition out of $Q$ (to $I$) is $q_1\lambda_q N_q$. Here, $1/\lambda_q $ is the quarantine duration, usually of the order of the incubation period \cite{Becker20,Linton20}. Similarly, if tested negative, the transition rate out of $Q$ into $S$ is $(1-q_1)\lambda_q N_q$. The total loss rate to ${\dot N}_q$ is $\lambda_q N_q$. 

Tracing those exposed to the infected, and testing to find if they are infected, are important steps in controlling the spread of the disease. Inherent in tracing such individuals are delays in tracing. Such delays cause increased transmission of the disease. If $p_t$ is the probability of tracing such individuals, then, $\alpha_t p_t N_s$ is the transition rate from $S$ to $T$. Subsequently, individuals testing positive will move to the infected compartment $I$ with a probability $q_2$ and the rest with a probability $(1-q_2)$ move to $S$. The total transition out of $T$ is equal to $\lambda_t N_{tr}$, where $1/\lambda_t $ is the time taken to trace the individuals. (There is also another possibility, namely, some individuals may show mild symptoms. Then, there would be a transition into $Q$. For the sake of simplicity, we have ignored this route.) Finally, the outward transitions from $I$ are the recovery and death rates respectively, $ \gamma_r N_i $ and $\kappa_d N_i$.

Collecting these terms, we have the following set of coupled nonlinear ordinary differential equations 
\begin{eqnarray}
\nonumber
\label{Sus}
\dot{N}_s &=& - (\alpha_s p_s + \alpha_s p_q + \alpha_t p_t) N_s -  p_i\beta_i N_i N_s \\
&+& (1-q_1)\lambda_q N_q + (1-q_2)\lambda_t N_{tr},\\
\nonumber
\label{Infec}
\dot{N}_i &=& \alpha_s p_s N_s +  p_{i}\beta_i N_i N_s \\
%\nonumber
 &+& q_1\lambda_q N_q + q _2 \lambda_t N_{tr}- (\gamma_r + \kappa_d) N_i,\\
\label{Quar}
\dot{N}_q &=& \alpha_s p_q N_s -\lambda_q N_q, \\
\label{Trac}
\dot{N}_{tr} &=& \alpha_t p_t N_s -\lambda_t N_{tr}, \\
\label{Rec}
\dot{N}_r &=& \gamma_r N_i,\\
\label{Dead}
\dot{N}_d &=& \kappa_d N_i.
\end{eqnarray}
(Here, we have suppressed $F(d_i)$ factor since it has been set equal to unity. Definitions of the notations are given in Table \ref{Select-Set-FM_N}). Note that the total infected population is given by $N_t= N_i + N_r + N_d$.

To begin with, we highlight a few features of the model equations. Our model, much as other compartment-type models, has several parameters. However, several of these are directly measurable and therefore can be obtained from the literature. A few others are related to testing protocols and again can be obtained from the literature or from relevant open sources \cite{WHO}. For instance, $\alpha_s p_s, \alpha_s p_q$ and $ \alpha_t p_t$ are directly related to testing rates and therefore, these are known for a given situation. A few other rate parameters such as $ \lambda_q, \lambda_t, \gamma_r$ and $\kappa_d$, are inversely related to measurable time-scales, such as the duration of quarantine $\tau_q$, time required for tracing $\tau_t$, time for recovery starting from illness $\tau_r$ and the time from illness to death $\tau_d$, respectively \cite{Linton20,Verity20}.

\begin{table}[h]
\begin{center}
\setstretch{1.3} 
\caption{\label{Select-Set-FM_N} Definitions of the notations for our calculation.}
\begin{tabular}{|c|c|}
\hline
{\bf Notation} & {\bf Definition}   \\
 \hline
$N_a(0) $ & Accessible population  \\
\hline
$\alpha_s$ & Testing rate per day per million \\
\hline
$p_s$ & Probability of testing positive\\
%\hline
%$ $ & Fraction  of the susceptible coming in contact with the infected & $0.1-0.5$ & $\cdots$\\
\hline
$ p_i$ & Probability of transmission of the disease \\
\hline
$ \beta_i \sim 1/N_a(0)$ & Transmission rate per contact  \\
\hline
$p_q$ & \multicolumn{1}{p{5.5cm}|}{\centering Probability of finding the \\ pre-symptomatic} \\
\hline
$\alpha_t$ & Tracing rate per day per million\\
\hline
$p_t$ & Probability of tracing \\
\hline
$ \lambda_q$ &   \multicolumn{1}{p{5.5cm}|}{\centering Rate  of quarantined individuals \\ testing positive}\\
\hline
$ \lambda_t$ &  \multicolumn{1}{p{5.5cm}|}{\centering Rate of tracing individuals exposed  \\ infected}\\
\hline
$ q_1$ & \multicolumn{1}{p{5.5cm}|}{\centering Probability of quarantined individuals \\ testing positive}\\
\hline
$ q_2$ & \multicolumn{1}{p{5.5cm}|}{\centering Probability of traced individuals \\ testing positive} \\
\hline
$ \gamma_r $ & Recovery rate of infected individuals \\
\hline
$ \kappa_d $ & Death rate of infected individuals \\
\hline
\end{tabular}
\end{center}
\end{table}

The present model includes two delay loops defined by Eqs. (\ref{Quar}) and (\ref{Trac}). These delays are natural to the implementation of the protocols. For instance, once quarantined, subsequent tests are conducted after quarantine duration to identify if quarantined individuals test positive or negative. Similarly, delays in tracing individuals are common. A more transparent way to describe these delay loops is through the integral representation of Eqs. (\ref{Quar}) and (\ref{Trac}), which {\it forms the definitions of the two populations $N_q$ and $N_{tr}$, respectively.} For instance, 
\begin{equation}
N_q(t) = \alpha_s p_s \int_{0}^ t dt' N_s(t') K(t-t'). 
\label{Nqint}
\end{equation}
When the kernel $K(t) $ is modeled using an exponential form with a single time scale $1/\lambda_q$, i.e., $K(t) = e^{- \lambda_q t}$, one can easily verify that differentiating Eq. (\ref{Nqint}) (using the Leibniz rule) leads to Eqs. (\ref{Quar}). The convoluted nature of the integral physically implies that those quarantined earlier will leave the quarantine sooner than those quarantined later.

Equations (\ref{Sus}-\ref{Dead}) constitute a set of coupled nonlinear differential equations. A standard procedure for further analysis of such equations is through numerical integration. While this is a necessary step, here we adopt a reductive analysis of the model equations by exploiting the fact that there are two main populations, namely, the susceptible (Eqs. \ref{Sus}) and the infected (Eq. \ref{Infec}). Furthermore, Eqs. (\ref{Rec},\ref{Dead}) are essentially decoupled from the rest (transitions to $R$ and $D$ are from $I$). These two features suggest that Eqs. (\ref{Sus},\ref{Infec}) can be decoupled from the rest of the equations. We refer to the decoupled equations as the reduced model equations. Since the two equations can be further reduced to a logistic-type equation (referred to as the reduced logistic equation), it can be analytically solved. As we shall see, analysis of this equation provides insights that prove to be useful for the analysis of the full model Eqs. (\ref{Sus}-\ref{Dead}). (We shall often refer to Eqs. (\ref{Sus}-\ref{Dead}) as "full model equations" to avoid confusion.)

\section{Concept of accessible population: The reduced model } 
\label{R-model}

 We now introduce the concept of 'accessible population for transmission of the disease'. To appreciate this concept, consider the spreading dynamics of a contagious disease in the absence of any interventions. Then, in principle, the entire population is exposed to the disease, and it may spread to the entire population (barring the possibility of the population acquiring herd immunity). In this case, the entire population is the accessible population. However, since no Government would like to see the entire population infected, interventional measures are enforced precisely to mitigate the risk of transmission and limit the population exposed to the disease to a minimum. In this case, the accessible population is expected to be a small fraction of the total population. These two limiting cases can be accommodated by hypothesizing the accessible population to be a fraction of the total population, where the fraction is determined by whether interventional measures are imposed or not. 

Consider dropping all terms except $\alpha_s p_s N_s$ and $  p_i \beta_i N_i N_s$ in Eqs. (\ref{Sus},\ref{Infec}). Then, these two equations get decoupled from the rest of the equations. Further, because all other inward/ outward transitions are removed, the character of the compartment $I$ changes from the active infected to the cumulative infected $I_t$ with $N_t$ denoting the corresponding population. Then, we have 
\begin{eqnarray}
\label{Acc}
\dot{N}_s &=& - \alpha_s p_s N_s -  p_{i}\beta_i N_t N_s,\\
\label{Cinfec}
\dot{N}_t &=& \alpha_s p_s N_s +  p_{i}\beta_i N_t N_s.
\end{eqnarray}
Noting that 
\begin{equation}
\frac{d}{dt}(N_s + N_t) = 0,
\end{equation}
 we have $ N_t + N_s =$ constant. Without loss of generality, we set $ N_t + N_s = N_s(0)$, the total population. 
Then, we get a single equation governing the cumulative infected population, given by 
\begin{eqnarray} 
\label{C-active}
{\dot N}_t &=& c + b N_t- a N_i^2,\\
\label{A}
a &= & p_i \beta_i, \\
\label{B}
b &= & p_i \beta_i N_s(0) -\alpha_s p_s,\\ 
\label{C}
 c &= & \alpha_s p_s N_s(0).
\end{eqnarray}
Equation (\ref{C-active}) has the well known form of logistic equation, extensively studied in the context of population dynamics \cite{Kingsland82}, with {\it a notable difference,} namely, the parameters $a,b$ and $c$ have a well defined interpretation as discussed above. We refer to Eq. (\ref{C-active}) as the reduced logistic equation. (For brevity we often refer to $\alpha_sp_s$ and $p_i\beta_i$ as testing and contact transmission rates, respectively.) 

We begin with a few observations on the relative magnitudes of the model parameters in the absence and the presence of interventions. Consider a situation when there are no constraints. Then, one should expect that the testing rate ($\alpha_s p_s$) to be low (compared to the lock-down period)  due to absence of any guidelines from policy makers. Similarly, since infected individuals carry on with their routine activity, the number of contact transmissions is high and hence, the contact transmission rate ($p_i \beta_i $) is expected to be high (compared to when interventions are in place). Then, the total accessible population denoted by $N_a(0)$ is the entire population of the region or the country, i.e., $N_a(0) = N_s(0)$. In contrast, when interventions are in place, testing rates are high to ensure identification of the infected, therefore, $\alpha_s p_s$ is high. In this situation, since the mobility of individuals is restricted, the number of contacts is severely limited, i.e., $  p_i \beta_i $ will be small. Therefore, the accessible population $N_a(0)$ is expected to be small compared to the total population $N_s(0)$. These two limiting cases can be written as  $N_a(0) \approx {\cal F} N_s(0)$. These qualitative statements about the accessible population will be made quantitative by carrying out a detailed analysis of Eq. (\ref{C-active}). 

Consider the initial growth of Eq. (\ref{C-active}) by dropping the quadratic term. Then, we have 
\begin{equation}
\frac{d}{dt} N_t = c + b N_t. 
\label{Linear}
\end{equation}
The solution is given by 
\begin{equation}
 N_t = \frac{c}{b}\big(e^{bt}-1\big)+ N_t(0) e^{bt},
\label{Linear-Ni}
\end{equation}
where $N_t(0) $ is the initial number of infections. As can be seen, the growth rate given by $b \approx  p_i \beta_iN_s(0)$ depends on $N_s(0)$, the total population. Therefore, the growth rate can be high. In addition, the pre-factor for the exponential growth term (in Eq. \ref{Linear-Ni}) depends not only on $ N_t(0)$ but also on $c/b=\alpha_s p_s/ p_i \beta_i$. Thus, the initial growth depends on relative magnitudes of $ N_t(0)$ and $\alpha_s p_s/ p_i \beta_i$.

It is straightforward to obtain the solution of Eq. (\ref{C-active}). (See Appendix for details.) Here, it is adequate to consider the solution in terms of the parameters $a,b,$ and $c$, given by 
\begin{equation}
N_t = \frac{\big(\frac{b}{a} N_i(0) + \frac{c}{a} \big) e ^{\, b t} + \frac{c}{a} + \frac{ac}{b}}{ (N_t(0) + \frac{ac}{b} \big) e ^{\, b t} - N_t(0) + \frac{b}{a} }.
\label{Full-solu} 
\end{equation}
We now examine two limiting cases. For short times, $ N_t$ tends to $ ( N_t(0) + \frac{c}{b} ) e ^{\, b t}$ (since the denominator is dominated by $b/a = N_s(0)$), consistent with the short time solution given by Eq. (\ref{Linear-Ni}). For long times however, $N_t$ tends to $b/a = N_s(0)$, the total population.

The self-limiting nature of Eq. (\ref{Full-solu}), a characteristic feature of logistic equations, is evident from the fact that $ N_t$ tends to $ N_s(0)$. In other words, the entire population becomes accessible for transmission of the disease. Clearly, the situation can only represent the growth of infection in the absence of any kind of interventions.

On the other hand, the effect of all interventions is to limit the contact transmission rate, thereby limiting the proportion of the exposed population to the disease to a small fraction. It is this that we call the accessible population. In other words, the accessible population $N_a(0)$ is of the same order as the infected population. This can be written as $ N_t \sim N_a(0) \approx {\cal F} N_s(0)$, where ${\cal F}$ is a small fraction. It must be noted here that the value of the fraction ${\cal F}$ depends on the nature of interventions in force. 

However, within the scope of the reduced logistic model, the evolution of $ N_t$ is independent of the values of the parameters $ \alpha_s p_s$ and $ p_i \beta_i$ during the absence or presence of interventions. As a consequence, the asymptotic value of the cumulative infected population is always ${N}_t=N_s(0)$, the entire population. Therefore, demonstrating the accessible population is a small fraction of the total population 
is outside the scope of Eq. (\ref{C-active}) and the full model Eqs. (\ref{Sus}- \ref{Dead}). An independent way of demonstrating $N_a(0) \approx {\cal F} N_s(0)$ is desirable.

\begin{figure}
\setstretch{1.3} 
\vbox{
\centering 
\includegraphics[height=4.5cm,width=8.50cm]{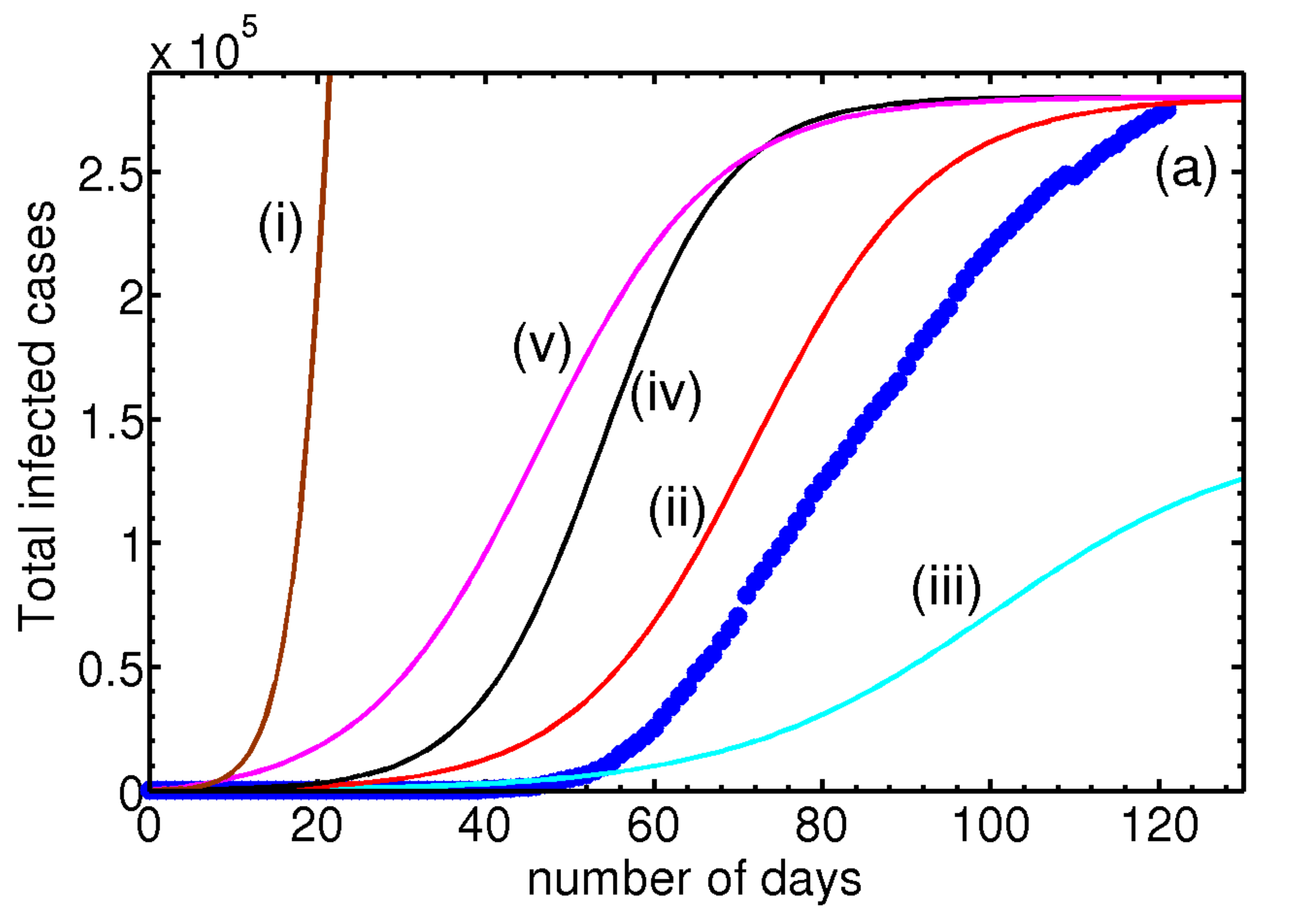}
\includegraphics[height=4.5cm,width=8.5cm]{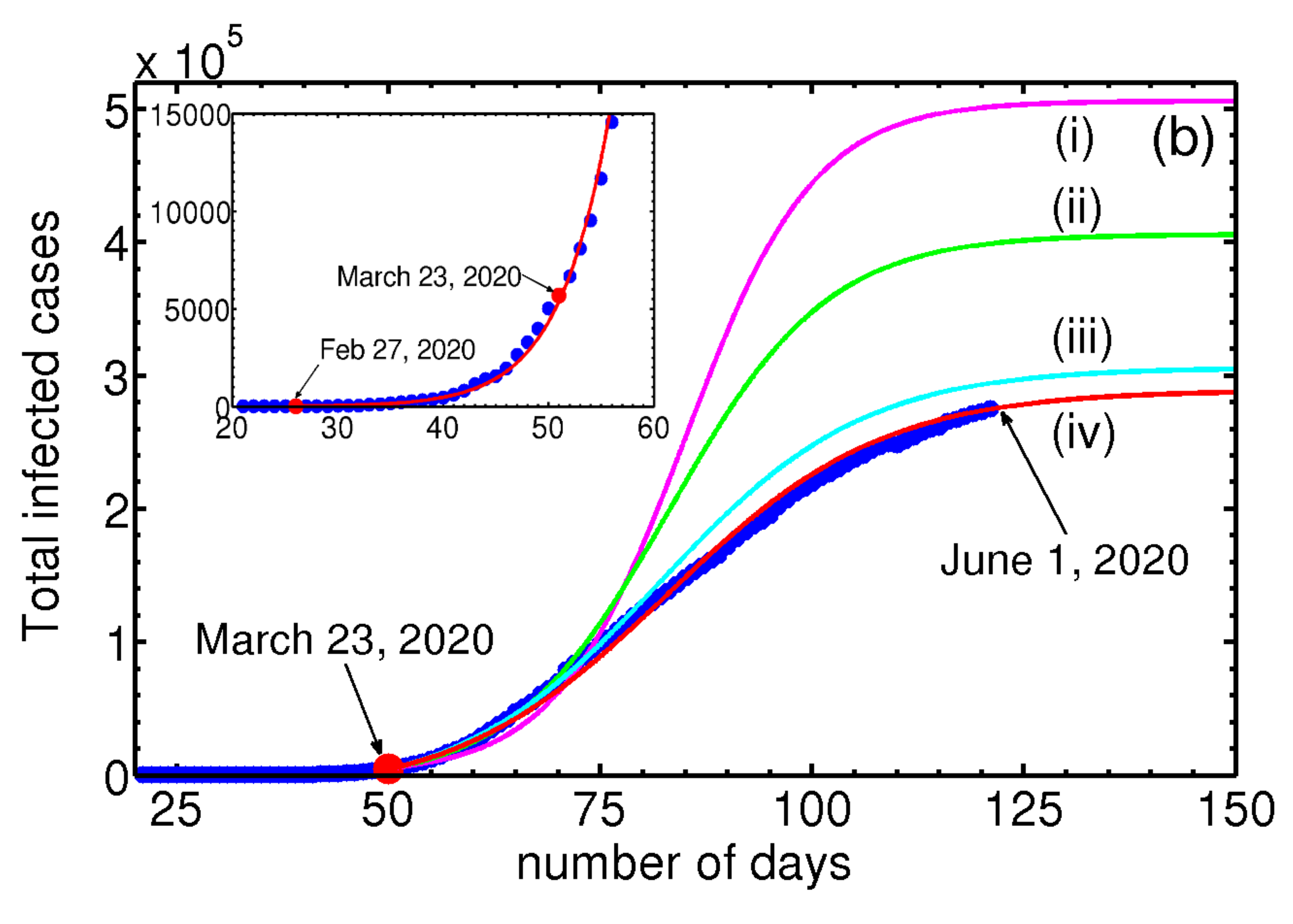}
\includegraphics[height=4.5cm,width=8.5cm]{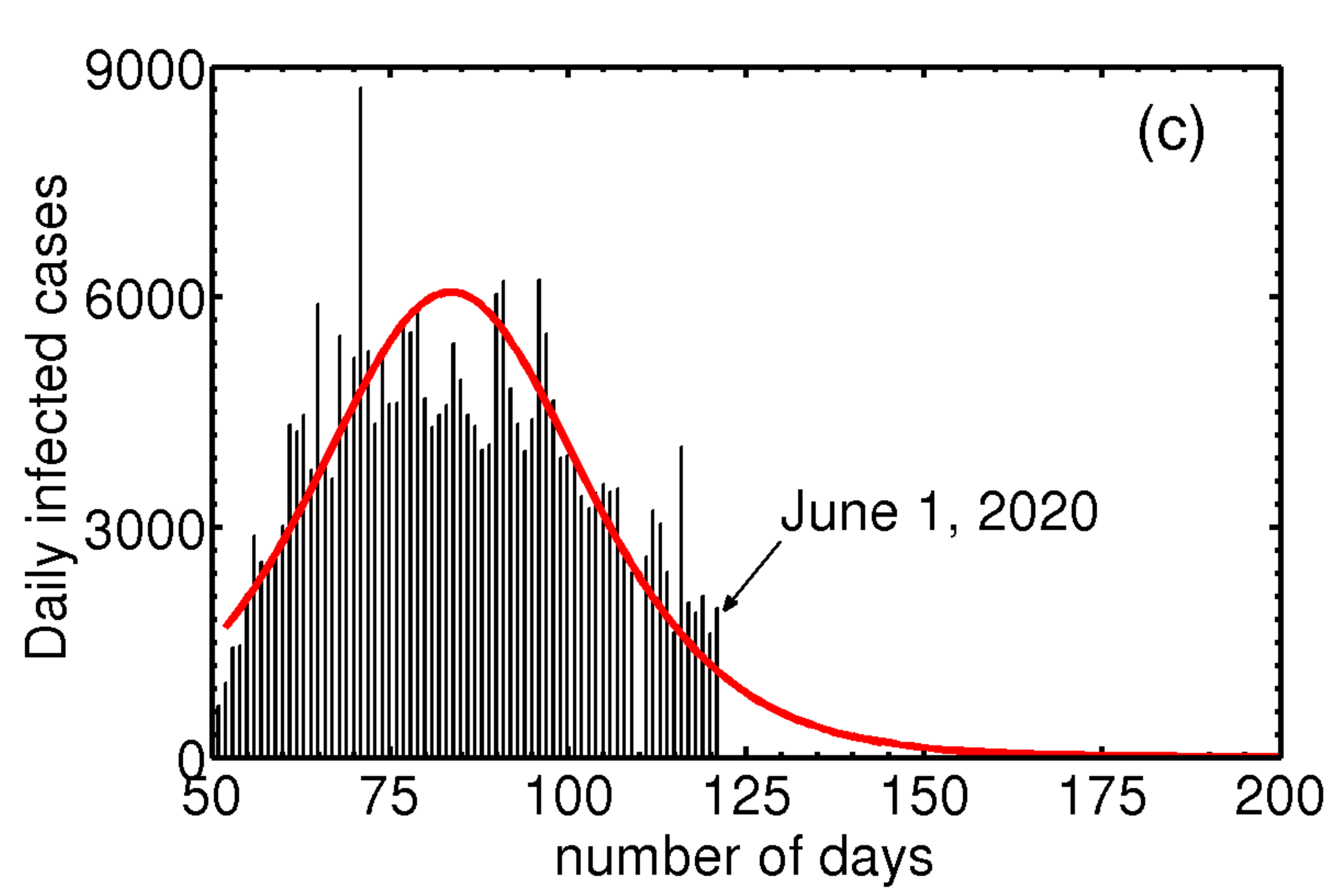}
}
\caption{(Color online) (a) Plots of $N_t$ for decreasing values of $N_a(0)$ as a function of time  : (i) $N_a(0) =8\times 10^5$, (ii) $2.8\times10^5$, and (iii) $1.45 \times 10^5$ respectively, keeping $ p_i\beta_i=3.3913\times 10^{-7}$ and $\alpha_s p_s= 1.0 \times 10^{-4}$ fixed. (ii) shows decreasing $N_a(0)$ by a factor of 5.51 leads to slow increase in $N_t$. (iv) Plot of $N_t$ for $ p_i\beta_i = 4.752 \times 10^{-7}$, keeping $N_a(0) = 2.8 \times 10^5$ and $\alpha_s p_s =1.0\times 10^{-4}$. Smaller values of $p_i\beta_i $ take longer time for $N_t$ to grow as is clear (see (ii) and (iv)). (v) Plot of $N_t$ for $\alpha_s p_s = 1.1 \times 10^{-3}$, keeping $N_a(0) = 2.8 \times 10^5$ and $ p_i\beta_i =3.3913 \times 10^{-7}$ fixed. Increase in $\alpha_s p_s$ leads to a faster initial growth seen in (v) and (iii). Also shown is the cumulative number of infections ($\bullet$) for the UK. (b) Figure shows the two-phase evolution of the disease. The inset shows the good fit using Eq. (\ref{Full-solu}) with the cumulative infected cases for the UK ($\bullet$) prior to March 23, 2020. Parameter used are $N_a(0)=3.27\times 10^5, N_t(0)=13, \alpha_s p_s=1.0 \times 10^{-5}, p_i\beta_i =6.4622\times 10^{-6}$. Post lock-down period: The four curves (i-iv) correspond to the four iterations of $N_a(0)$ values. (i) $N_a(0)=5.0 \times 10^5,\alpha_s p_s=0.0$, (ii) $N_a(0)=4.0\times 10^5,\alpha_s p_s=1.8\times 10^{-3}$, (iii) $N_a(0)=3.0 \times 10^5$, and (iv) $N_a(0)=2.8 \times 10^5,$ with $\alpha_s p_s= 4.83\times 10^{-3}$ for latter two. The initial value of $N_t(0)=5687$ on March 23, 2020. (c) Plots of daily new cases for the UK and the calculated daily new cases using the reduced logistic model predicted.}
\label{UK-Fit-logistic}
\end{figure}

\subsection{Quantitative estimate of the accessible population}
\label{Accpop}

Since the factor ${\cal F}$ is not well determined, there is a necessity to get a better estimate of this parameter or the accessible population $N_a(0)$. Assuming that the accessible population is of the order of $N_t$, we assume that $N_a(0) \approx {\cal F}N_s(0)$. This is equivalent to using $N_a(0)$ in place of $N_s(0)$ in Eq. (\ref{Full-solu}). Then we   numerically evaluate the dependence of $ N_t$ on the parameters on $N_a(0), \alpha_s p_s$, and $  p_i \beta_i $. Given the fact that the disease evolves, we expect that the accessible population 
$N_a(0)$ also evolves with time and in the early stages of evolution, $N_a(0)$ will be small, even in the absence of interventions.

Consider the dependence of $N_t$ on $N_a(0)$, keeping $\alpha_s p_s$ and $  p_i \beta_i $ fixed. 
We find that even for relatively large values $N_a(0)$, $N_t$ grows exponentially; for intermediate values, a near saturation value is reached in a relatively short duration of 10-15 days; and for small values, the saturation value is not reached even after 100 days. These features are illustrated in Fig. \ref{UK-Fit-logistic}(a) in plots (i-iii) for $N_a(0)= 8 \times 10^5, 2.8 \times 10^5$ and $N_a(0) = 1.45\times 10^5$ respectively, keeping $ p_i\beta_i =3.3913 \times 10^{-7}$ and $\alpha_s p_s= 1.0 \times 10^{-4}$. We have also examined the influence of $p_i\beta_ i$, keeping $N_a(0) = 2.8 \times 10^5$ and $\alpha_s p_s =1.1 \times 10^{-3}$. Smaller values of $p_i\beta_i$, $N_i$ take  longer time for the infection (${N}_t$) to grow. This feature can be seen from the curves (iv) and (iii) for $p_i\beta_i = 4.7522 \times 10^{-7}$ and $ 3.3913 \times 10^{-7}$ respectively. We have also examined the growth dependence of ${N}_t$ on $\alpha_s p_s$, keeping the other two parameters fixed. The dependence of ${N}_t$ on this parameter is similar to that on $p_i\beta_i $. The curve (v) taken together with (ii) shows that increasing $\alpha_s p_s$ also leads to faster initial growth of ${N}_t$. In the same plot, we have also plotted the total number of infected cases $(\bullet$) for the UK.

A careful scrutiny of the total coronavirus cases ($\bullet$) in the UK shows that it is similar- both in magnitude and shape -to the plot of $N_t$ corresponding to $N_a(0) = 0.28 \times 10^6$ marked (ii) shown in Fig. (\ref{UK-Fit-logistic}a). This similarity suggests two important points. First, noting that the UK is under lock-down, one expects that the accessible population is a small fraction of the total population, and therefore we see that {\it the order of magnitude of the accessible population $N_a(0)$ used is comparable to that of the infected population $N_t$} shown in curve (ii). The figure also shows that as much as all populations dynamically evolve during the development of the pandemic, $N_a(0)$ also keeps evolving with time. Second, the similarity in shape of the UK data ($\bullet$) with the sigmoidal shape of the logistic solution raises a question whether the similarity is accidental. If not, can this be used to fit the UK data? 
 
\subsection{Data Assimilation and fitting}
\label{Logistic-fit}

However, considering the complex dynamics of the highly contagious virus and the fact that logistic equation can at best represent simple situations, any attempt to fit the data appears ambitious. Even so, it is tempting to examine if Eq. (\ref{Full-solu}) could be used to fit the coronavirus data for some country/region. To do this, we first note that the reduced model equation contains just three parameters and the dependence of $N_t$ on these parameters has already been examined [see Fig. \ref{UK-Fit-logistic}(a)].

In most countries, the development of the disease falls into two phases, namely, the initial period when Governmental constraints are absent, referred to as phase one and the period beyond the lock-down date, called phase two. In the case of the UK, the first case was reported on January 31, 2020. Subsequently, the lock-down was imposed on March 23. Thus, we need to fit the data for the period January 31 to March 23 and then the rest.

Consider the period between January 31 and March 23, 2020. Briefly, the fitting procedure adopted here is to equate the initial growth rate of infections obtained from the coronavirus data with the model growth rate given by Eq. (\ref{Linear-Ni}) (or Eq. \ref{Full-solu}). Using the fact that the accessible population is of the order of the total number of infections, we use a trial value of $N_a(0)$ (assumed to be a few times larger than the infected population) to fix the parameter $\beta_i$. Then, the correct value of $N_a(0)$ that provides the best fit for the entire data is found iteratively by decreasing $N_a(0)$ so as to fit an increasing number of data points. The procedure is illustrated below.

 Here, we use the analytical solution given by Eq. (\ref{Full-solu}) (or solving Eqs. \ref{Acc}-\ref{Cinfec}) with parameters and initial conditions appropriate for the unconstrained growth. 
Recall that the testing rate parameter $\alpha_s p_s$ is low during the initial period and the contact transmission rate parameter $ p_i\beta_i $ would be high. The values of these two parameters in the lock-down period are just the opposite. 

Consider the first phase where virus transmission is unconstrained. A careful perusal of the UK data shows that a smooth increase in the infected numbers starts on Feb. 26, 2020, when the number infected stood at ${N}_t = 13$. The local growth rate obtained from the data over 8 days was found to be 0.25849/day. Equating this with the model growth rate given by $ p_i \beta_i N_a(0)$ (in Eq. \ref{Linear-Ni}), with a trial value of $N_a (0) = 4.0 \times 10^{5}$ fixes a value of $\beta_i = 6.4622 \times 10^{-6}$. The solution of Eq. (\ref{Full-solu}) (or Eqs. \ref{Acc}-\ref{Cinfec}) obtained using the initial condition ${ N}_t = 13$ keeping $\alpha_s p_s= 0$, passes through several more data points than 8. In the next iterations, we reduce $N_a(0)$, keeping in mind that the solution should pass through a larger number of data points. In addition, since the initial growth rate (Eq. \ref{Linear-Ni}) depends also on $c/b = \alpha_s p_s/ p_i \beta_i $, a proper value of $\alpha_s p_s$ is required for a good fit. We find that just one iteration of reducing $N_a(0)$ to $N_a(0)= 3.27 \times 10^5 $ with $\alpha_sp_s = 1 \times 10^{-5}$ fits the data well for the period from Feb. 27 to March 23, 2020, as shown in the inset of Fig. \ref{UK-Fit-logistic}(b).

Fitting the data for the second phase follows the same iterative procedure except that the number of iterations is greater for the second phase due to the large number of data points. The number of infections as on March 23 stood at ${ N}_t = 5687$. This number matches with the predicted value of $N_t$ as on March 23, 2020, obtained from Eq. (\ref{Full-solu}) for the first phase. (See the inset in Fig. \ref{UK-Fit-logistic}(b)). Since the effect of lock-down is expected to take some time to manifest, we have taken the local slope over 17 points from the lock-down day is 0.13/day. This slope is equated with model growth rate using a trial value of $N_a(0) = 5.0 \times 10^5$ ($\alpha_s p_s = 0$) to obtain $\beta_i = 2.602 \times 10^{-6}$. Using the initial condition ${ N}_t = 5687$ in Eq. (\ref{Full-solu}) (or solving Eqs. \ref{Acc}-\ref{Cinfec}), we find that the solution (i) (with $ \alpha_s p_s = 0$) passes through a few more than 17 points. In the next iterations, we reduce $N_a(0)=4.0\times 10^5$ and compute the solution taking into account the contribution from $\alpha_s p_s=1.8\times 10^{-3}$. The solution (ii) passes through several more data points. Two further iterations for successively smaller values of $N_a(0)=3.00 \times 10^5$ and $N_a(0)=2.9 \times 10^5$ are used to obtain the solution marked (iii) and (iv), respectively. ($\alpha_s p_s =4.83\times 10^{-3}$ is used in both cases.) This is shown in Fig. \ref{UK-Fit-logistic}(b). As is clear from the Fig. \ref{UK-Fit-logistic}(b), solutions (ii) and (iii) are seen to pass through successively larger number of points. Surprisingly, the solution curve labeled (iv) with $N_a(0) = 2.9\times 10^5$ fits the entire data fairly well. Note the increasing trend of the values of $\alpha_s p_s$ for successive iterations. This feature is consistent with the steadily increasing testing rates routinely used for proper enforcement of lock-down. This feature is easily incorporated by parameterizing $\alpha_s p_s$ with time.

 Now consider the  estimation of the end time of the epidemic, usually  defined as  the time when no new infected cases are reported. Noting the close fit of the model predicted cumulative infected population $N_t$ with the UK data  seen in Fig. \ref{UK-Fit-logistic}(a), $\frac{dN_t}{dt}$ gives the model calculated new infected cases. This can be compared with the UK data for the daily new cases. This is  shown in Fig. \ref{UK-Fit-logistic}(c). As can be seen from the figure the general profile of the model predicted daily new infected cases  matches  well with the published data for the UK. The estimated end time of the epidemic to be last week of June.

Apart from providing a close fit for the entire data, the method appears to have predictive power, as is clear from the curve (iv), which shows that the rate of slowing of the total number of infections is decreasing. The predicted saturation value is $ \sim 2.9 \times 10^5$. A near-saturation value is likely to be seen by the second week of June. These results suggest that the reduced logistic model can be used for obtaining  fit for the COVID-19 data for other countries as well. The close fit in itself  is attributable to the fact that the dominant contribution to the growth of the total infected population $N_t$ comes from the two direct transitions. On the other hand, Eq. (\ref{Full-solu}) does not include outward transitions (the recovered and the dead), and also the inward quarantine and tracing transitions. Therefore, the estimated saturation value and the projected future development should be taken with some reservations. This will be clear once the full model  equations are analyzed and a fit with COVID-19 data for the UK is accomplished. Despite these limitations, because the reduced logistic equation retains basic growth contributions to the cumulative infected $N_t$, the fit with the data appears reasonable.

There are attempts to use logistic equations to get insights into the dynamics of COVID-19 transmission \cite{Thyaga20,Kriston20}. For instance, a five-parameter hierarchical logistic model has been used to fit the observed data to project the cumulative number of cases for several countries \cite{Kriston20}. The parameters entering in the model are determined by the fitting procedure.

\section{The full model} 

One of the challenges of compartmental models is the difficulty associated in making accurate predictions, mainly attributable to the uncertainties in obtaining proper estimates of the parameters \cite{Dehning20,Salje20,Friston20,Linton20,Verity20}. For the same reason, forecasting is even more challenging. Often, several factors may also contribute to the same parameter, making it difficult for proper interpretation. In our model, however, several parameters in Eqs. (\ref{Sus}- \ref{Dead}) are related to measurable quantities. For instance, the parameters $\alpha_s p_s$, $\alpha_s p_q$ and $\alpha_t p_t$ respectively represent rates of testing positive, rates identified as pre-symptomatic, and tracing rate of those exposed to the infected. 
 Similarly, parameters $\lambda_q, \lambda_t,\gamma_r$ and $\kappa_d$ are inversely related to quarantine duration $\tau_q = 1/\lambda_q$ and time required for tracing $\tau_t = 1/\lambda_t$, time from illness to recovery $\tau_r=1/\gamma_r$, and time from illness to death $\tau_d=1/\kappa_ d$. Though these quantities are country/region-specific, their values have been estimated in the literature \cite{Tang20,Tang20a,Zhao20,Linton20,Verity20,Gatto20,Jung20,Dorigatt20,Anasta20,Wang20}. Some values are also available in the public domain \cite{WHO,ECDP}. One parameter that is hard to estimate is the contact transmission rate $\beta_i$, which is already estimated in the context of the reduced logistic equation. 

\begin{table}[h]
\setstretch{1.3} 
% table caption is above the table
\caption{\label{Select-Set-FM} Post lock-down period: Select set of parameter values serving as a reference set. $q_1=q_2=0.08$ and $ p_i = 0.1$.
}     % Give a unique label
% For LaTeX tables use
\begin{tabular}{llll}
\hline\noalign{\smallskip}
$\alpha_s p_s$ & $\alpha_s p_q$ & $\alpha_t p_t$ & $  p_i\beta_i$   \\
\noalign{\smallskip}\hline\noalign{\smallskip}
$9.0\times 10^{-4}$ & $3.64 \times 10^{-2}$ & $3.6 \times 10^{-3}$ & $ 3.4693\times 10^{-7} $\\
%\noalign{\smallskip}\hline\noalign{\smallskip}
%$0.08$ & $0.08$ & $1/14$ & $1/3$   \\
\noalign{\smallskip}\hline\noalign{\smallskip}
$\lambda_q$ & $\lambda_t $ & $\gamma_r$ & $\kappa_d$  \\
\noalign{\smallskip}\hline\noalign{\smallskip}
$1/14$ & $1/3$ &$1/42$ & $1/70$  \\
\noalign{\smallskip}\hline
\end{tabular}
\end{table}

\begin{figure}
\setstretch{1.3} 
\vbox{
\centering 
\includegraphics[height=6.0cm,width=8.50cm]{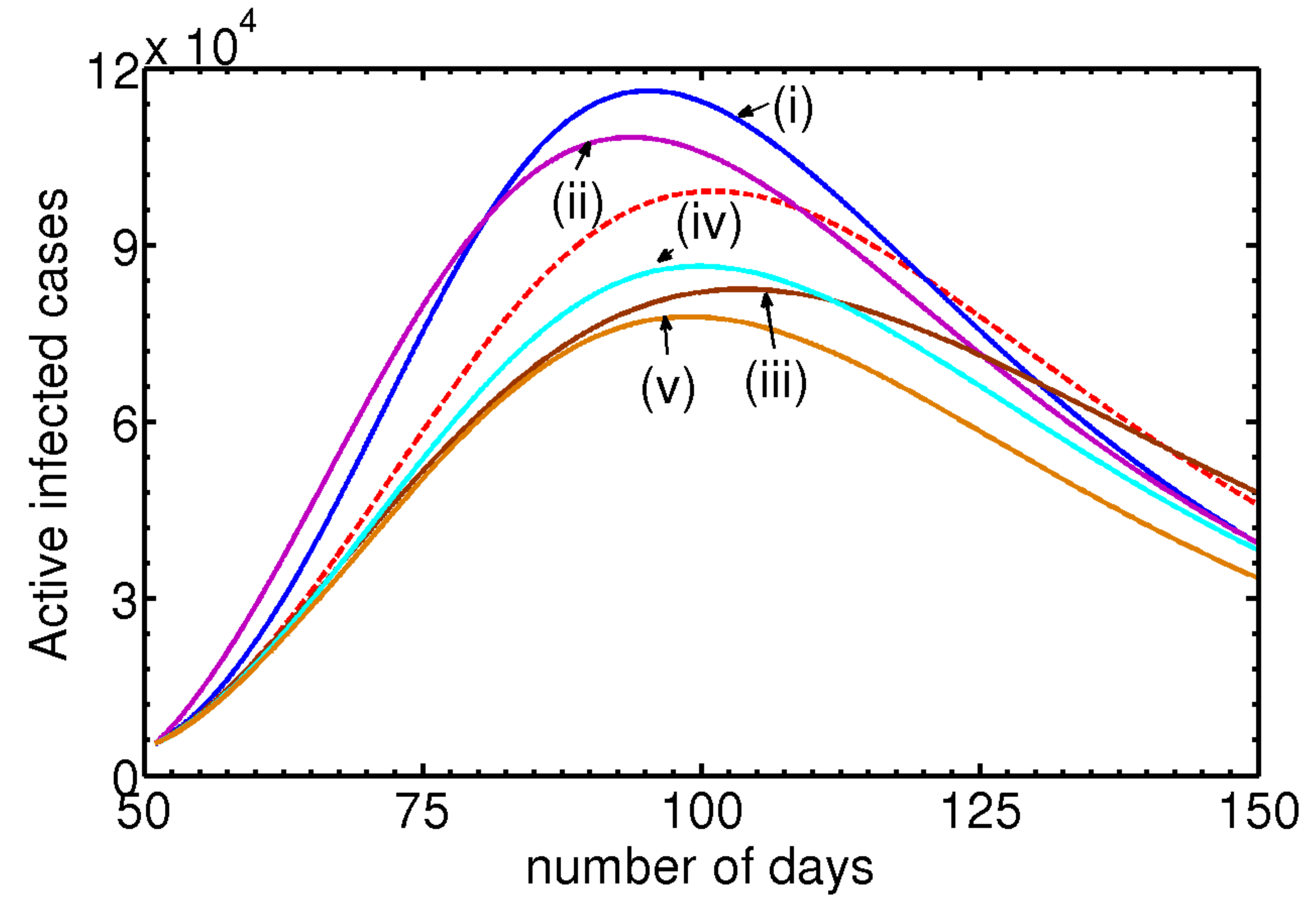}
}
\caption{(Color online) Calibration of parameters for identifying the relative importance of transitions contributing to $N_i$ by varying one parameter, keeping all others parameters fixed at reference values listed in Table \ref{Select-Set-FM}. The dotted curve is the reference plot for the active infected population $N_i$ corresponding to the values in the Table \ref{Select-Set-FM}. (i) Plot of $N_i$ for a 50\% increase in $ p_i\beta_i $ showing a substantial increase in the peak height (38\%) and a shift by 25 days. (ii) Similar effects  are seen when $\alpha_s p_s$ is increased by a factor 8. (iii) The plot shows a substantial decrease in the peak height with a marginal shift in its position when $\alpha_s p_q$ is increased by a factor of 2. (iv) The plot shows a decrease in the peak height as $\kappa_d$ is increased by 60$\%$. (v) Similar effect is seen when the recovery rate $\gamma_r$ is increased by 60\%.}
\label{Para-vary-NF}
\end{figure}

\subsection{Calibration of relative strengths of the parameters: insights into disease evolution}
\label{Calib}

However, the dynamical evolution of a nonlinear coupled set of equations such as Eqs. (\ref{Sus}- \ref{Dead}) is necessarily complex. Therefore, in the absence of appropriate values relevant {\it for the country/region}, a systematic method of finding optimized values of parameters that fit the data under consideration requires calibration of all parameters in the model. This will also help us to delineate the different time-scales participating in Eqs. (\ref{Sus}- \ref{Dead}). For instance, several of these parameters represent the growth or decay rates  of these populations. Our model contains eight time-scales which are inversely related to the transition rates. These are $\beta_i$, $\alpha_s p_s$, $\alpha_s p_q$, $\alpha_t p_t$, $\lambda_q$, $\lambda_t$, $\gamma_r$ and $\kappa_d$. These time scales control how these populations evolve with time. From the structure of Eqs. (\ref{Sus}-\ref{Dead}), it is clear that $N_i,N_q$ and $N_{tr}$ exhibit peaks as a function of time (days) whereas $N_r$ and $N_d$ grow monotonically. However, at what point of time do the peaks appear in these populations with the progression of the disease cannot be easily  determined since these are coupled nonlinear differential equations where the evolution of any  population depends on the evolution of all other populations \cite{Strogatz}. More importantly, if one is interested in fitting the model predicted growth of populations, which convey the disease status (such as the total infected, active, recovered and dead), estimating  the relative proportion of the populations as the disease evolves is necessary. Further, the total infected population commonly used to convey the disease status in daily briefings has  contributions from all populations. Therefore, delineating and  determining at what points of time each of these populations contribute to the total populations would provide required insight into further analysis.

Following the recently developed method in the area of plasticity \cite{Srikanth17,GA18,GA19}, we investigate the influence of the parameters to identify the relative importance of the transition rates. Since it is a multi-parameter space, we vary each parameter, keeping all other parameters fixed at the reference set of values listed in Table \ref{Select-Set-FM}. The results are illustrated using plots of the active infected population $N_i$. The dotted curve shown in Fig. \ref{Para-vary-NF} is the reference curve corresponding to the reference set of parameters given in Table \ref{Select-Set-FM}. As in the case of the reduced logistic model, the growth of $N_i$ depends sensitively on the contact transmission parameter $ p_i \beta_i$. (Note that in our model, this is the  only parameter that directly contributes to growth of infections.) A 50\% increase in the parameter induces a substantial increase (38\%) in the peak height with the position shifting towards earlier time (by 25 days) as is clear from the curve (i). Noting that the position of the peak, i.e., the turning point of $N_i$ is indicative of slowing down of the rate of infection, increasing peak height of $N_i$ suggests slowing down of infections occurs at higher values of infections, where as a shift towards shorter times implies that slowing down occurs earlier. A similar effect but of lesser magnitude (20\%) is seen when testing rate $\alpha_s p_s$ is increased by a factor 8 as is clear from (ii). 

  In contrast, an increase in the quarantining rate $ \alpha_s p_q$ by a factor 2 decreases the peak height comparable in magnitude to that induced by $\beta_i$ with  a shift in the peak position away from the origin by 20 days (see iii). A similar effect, but of lesser magnitude, is seen when $\alpha_t p_t$ is increased (not shown). The reduction in  the peak height of $N_i$ is understandable because the total inward transition is  $\alpha_s p_s + \alpha_s p_q +\alpha_t p_t $, and therefore increasing one of these  changes the relative weights. Physically, the decrease in the peak height (commonly referred to as 'flattening the curve' ) accompanied by a shift in the peak position for longer times implies that as more individuals are quarantined, the  disease control is facilitated. 
	
	We have also investigated the dependence of the recovery ($\gamma_r$) and death rate ($\kappa_d$) parameters on $N_i$. An increase in  $\gamma_r$ by 60$\%$ decreases the peak height substantially as is clear from (iv). Similarly, increase in the death rate $\kappa_d$ leads to a decrease in the peak heights of $N_i$ as is clear from (v). Clearly, increased recovery rate, a desirable feature, also leads to a decrease in the number of active infections. This feature  is also intuitively obvious. On the other hand, increased death rate, though not desirable, also  leads  to decrease in $N_i$. We have also investigated the influence of other parameters and find that $N_i$ is relatively insensitive to these parameters. Noting that any change in the parameter values relative to those corresponding to the reference curve changes the peak position and height, {\it we conclude that the parameters listed in Table \ref{Select-Set-FM} are close to the optimized values. }

The above analysis on the relative importance of the transition rates (equivalently the corresponding parameters), and the accompanying discussions, can now be used to delineate the contributions from different time scales participating in the transmission dynamics of the virus. As discussed above, the direct transitions from $S$ to $I$, namely, $p_i \beta_i$, to a lesser extent $\alpha_s p_s$, control the initial growth of $N_i$. The same also  to the initial growth of the total infected population $N_t$. Similarly, the delayed routes, namely, the quarantine $S$ and tracing $T$ contribute to the mid region of the evolution of $N_i$ and to  therefore to the total population $N_t$ also. Recall that the turning point of $N_i$ is controlled by the  balance between all inward transitions (from $S$ to $I$, and $Q$ and $T$ to $I$) and outward transitions ($I$ to $R$ and $D$ compartments). Then, the peak position of $N_i$ can be identified with the inflection point of $N_t$. Therefore, the time development beyond the point of inflection of $N_t$ is controlled by a balance of all inward and outward transitions. The insights from the above analysis identify three distinct stages in the evolutionary period of the disease, namely, the initial growth period, the mid developmental period and the final approach to saturation. As we shall see, this identification will be helpful in obtaining a good fit to the UK data.

\begin{figure}%[!h]
\setstretch{1.3} 
\vbox{
\centering
\includegraphics[height=6.20cm,width=8.5cm]{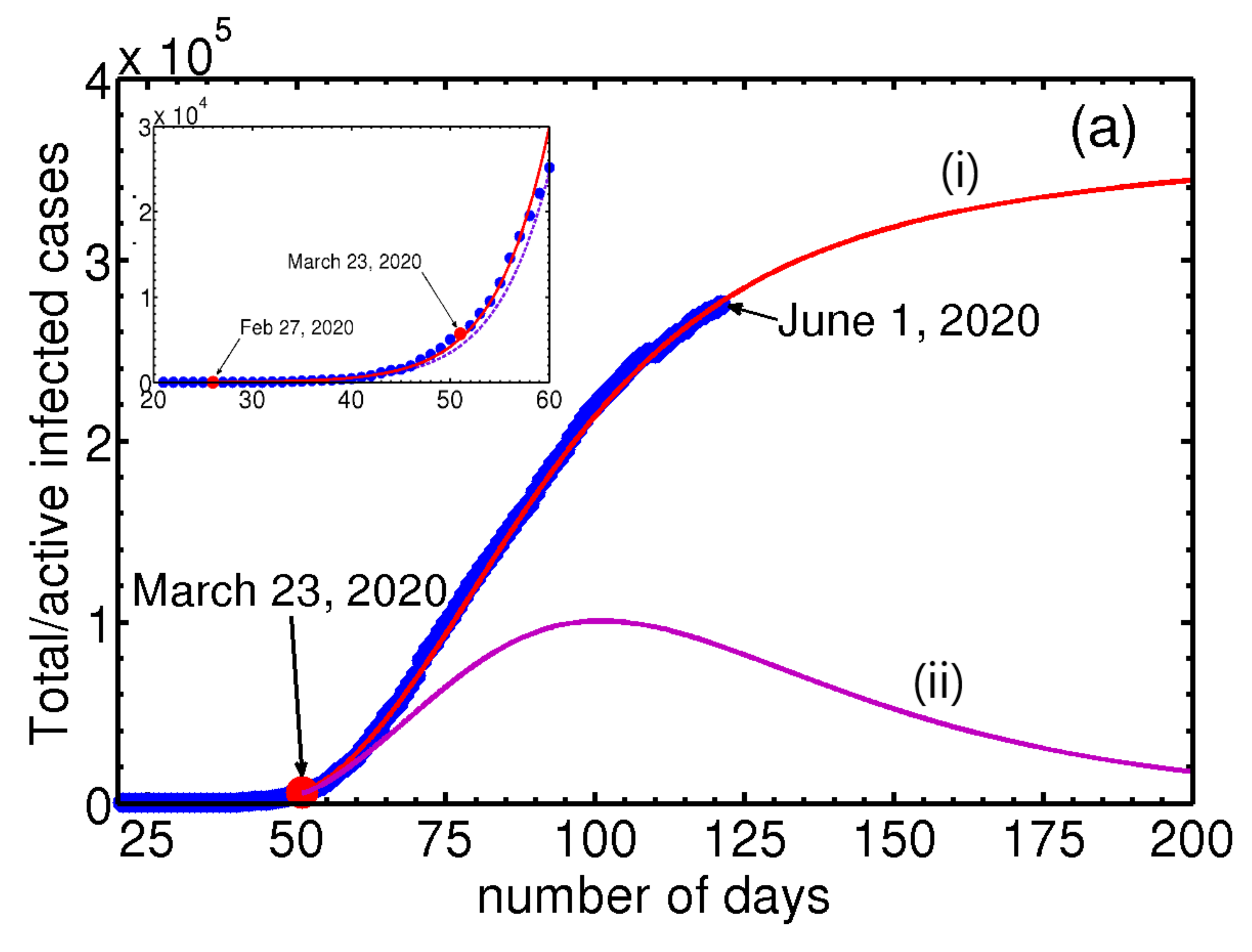}
\includegraphics[height=6.20cm,width=8.5cm]{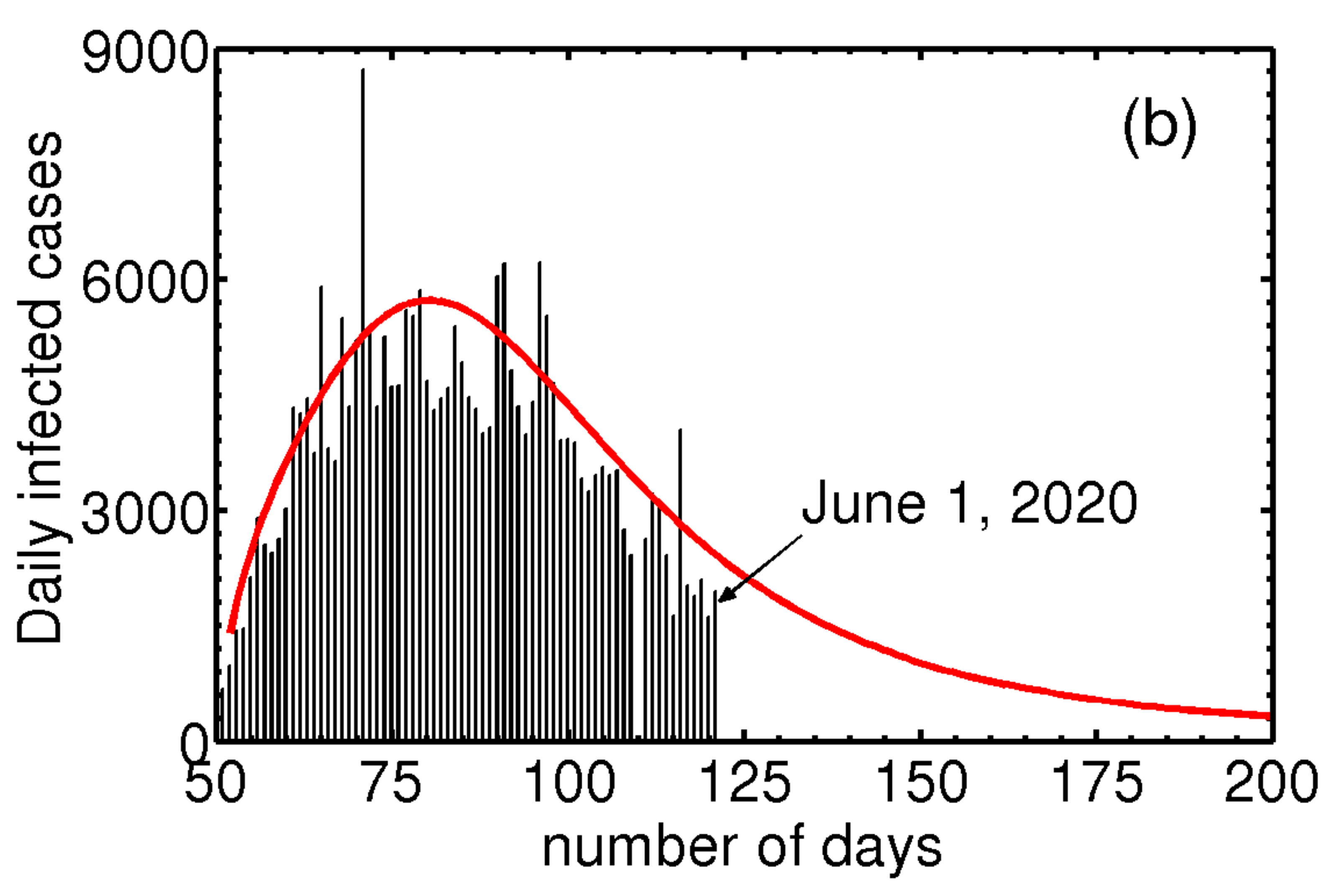}
\includegraphics[height=6.20cm,width=8.8cm]{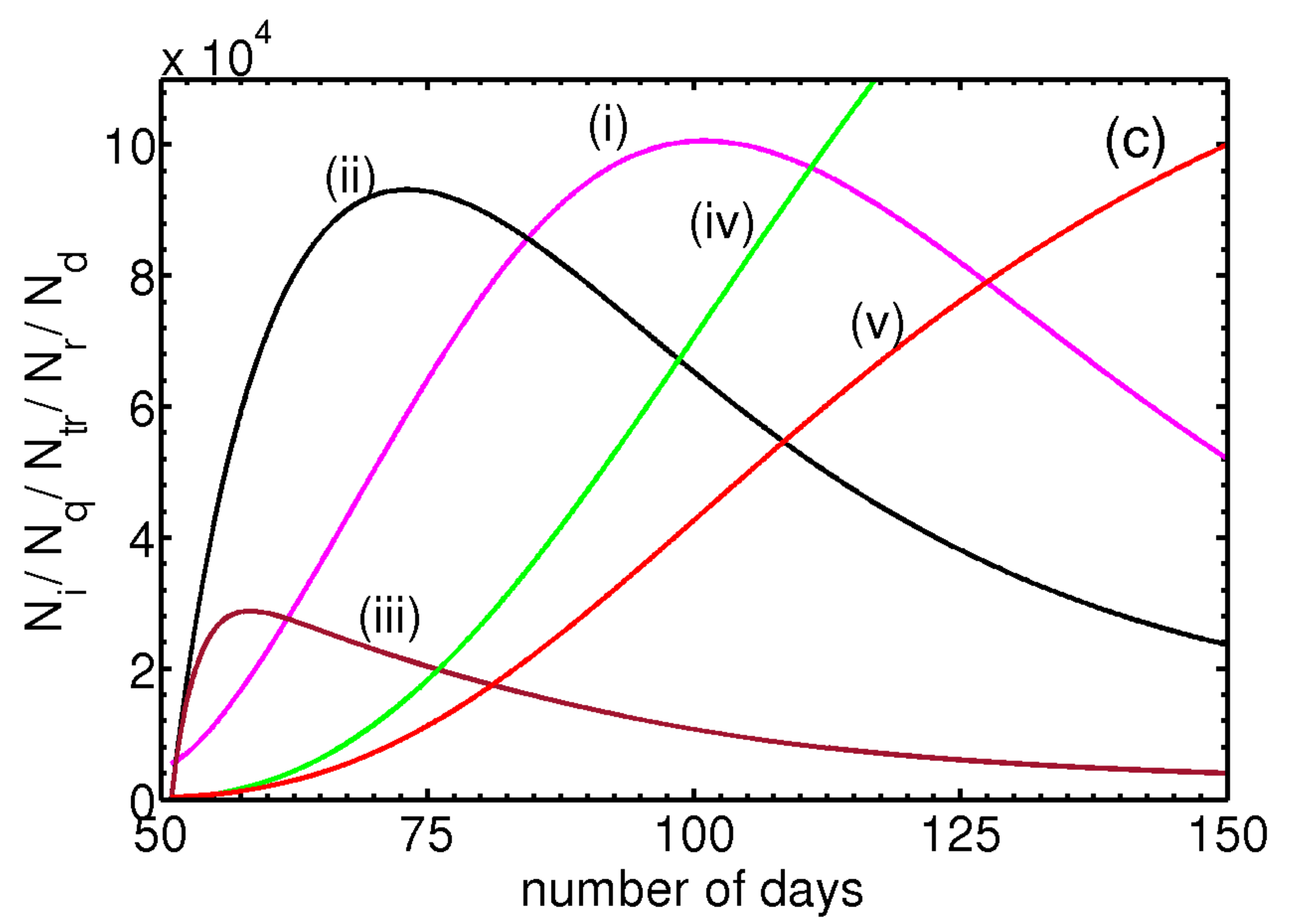}
}
\caption{(a) Inset: Plot of the total infected population (continuous curve), along with the UK data ($\bullet$) from February 27, 2020 till March 31. Also shown is the active infected $N_i$ (dotted curve). Post Lock-down period: Plots of the total infected population (curve marked i) along with the corresponding data for the UK ($\bullet$) from March 23, 2020. Curve marked (ii) shows the active infected population. (b) Plots of the reported daily new cases for the UK along with the model predicted daily new cases. (c) Plots of the infected (i), quarantined (ii), traced (iii), recovered (iv) and deceased (v), starting from March 23, 2020. The values of the parameters used are given in Table \ref{Select-Set-FM}. 
 } 
\label{UK-Fullmodel}
\end{figure}

\subsection{Data assimilation and forecast}

Having demonstrated that the two direct transition rates $ p_i \beta_i$ and $\alpha_s p_s$ are the dominant contributions to the growth of $N_i$ and having assessed the relative importance of other transitions, we now consider the solution of the full model Eqs. (\ref{Sus}- \ref{Dead}) with a view to obtaining the best possible fit with the COVID-19 United Kingdom data. Attempt will also be made to forecast the future progression of the disease.

Recall that the spread of coronavirus in the UK falls into two phases of development. During the first phase prior to  March 23, 2020, there were no constraints and the disease transmission was free. After the lock-down date, 
the transmission is restricted. Therefore, the model parameters and the initial conditions relevant for the two phases are different. As in the reduced model, we assume that the dynamics of the disease transmission is limited by the accessible population $N_a(0)$ and not by the total population $N_s(0)$, i.e., $N_i \sim N_a(0) \approx {\cal F} N_s(0)$. Note that both $N_i$ and therefore $N_a(0)$ depend on whether there are any interventions or not, and  also on the type of interventions.

Consider the period between January 31 and March 23 corresponding to the initial phase. 
Here, we note that  prior to the lock-down date, there would be no quarantining and tracing procedures and therefore, it is adequate to solve Eqs. (\ref{Sus},\ref{Infec},\ref{Rec},\ref{Dead}), More specifically, we ignore the delayed inward transitions into $I$. Furthermore, in the first few days of the development of the disease, we may assume that the total number of the infected cases $N_t$ is equal to the active infections $N_i$. Finally, since, we plan to fit the model solution with the UK data \cite{WHO}, publicly available coronavirus data for the total number infected, active infected, recovered and the dead are useful in further optimizing the parameters. Unfortunately however, only the total numbers of the infected and the dead are made available in the UK.

Now we are in a position to solve the relevant equations for the first phase. As discussed in Section \ref{Logistic-fit}, we use February 27, 2020 as the starting day for the first phase evolution of Eqs. (\ref{Sus},\ref{Infec},\ref{Rec},\ref{Dead}). The local growth rate of 0.25849/day (on the starting day) over 8 days obtained from the log-linear plot of the cumulative infected cases for the UK is equated with the model growth rate given by $ p_i \beta_i N_a(0)$ to fix $\beta_i= 6.4622 \times 10^{-6}$ by using the initial value of $N_a(0) = 4.0 \times 10^5$. Further, using the initial conditions for $N_t(0)=N_i(0) =13, N_r(0) = 0$ and $N_d(0) =0$, we solve Eqs. (\ref{Sus},\ref{Infec},\ref{Rec},\ref{Dead}) from February 27 to March 23 by choosing a value for $\alpha_s p_s $ that gives the best fit to the data for the period. (Here, $\alpha_s p_s = 3.6 \times 10^{-6}$ and the values of other relevant parameters are those listed in Table \ref{Select-Set-FM}.) The model-predicted total infected population $N_t$ (continuous curve) along with the data points ($\bullet$) is shown in the inset of Fig. \ref{UK-Fullmodel}(a). Clearly, the match is seen to be very good. Also shown is a plot of active infections $N_i$ (dotted curve). Equations (\ref{Sus},\ref{Infec},\ref{Rec},\ref{Dead}) also provide the values of $N_i,N_r$ and $N_d$ on March 23, 2020. These are $N_i= 5407, N_r = 400, N_d = 285$.

Now we consider the solution of Eqs. (\ref{Sus}- \ref{Dead}) with a view to obtaining the best fit for the UK data for the period starting from March 23, 2020. Here again, we first find the growth rate from the data and equate it with the model growth rate. Here, we note that the effect of lock-down  is expected to manifest after some time. For this reason, we use a 17-point slope in the log-linear plot. Equating 0.13/day with $ p_i \beta_i N_a(0)$ and using $N_a(0) = 3.75\times 10^5$ we get $ \beta_i = 3.4693\times 10^{-6}$. The initial values used for evolving Eqs. (\ref{Sus}- \ref{Dead}) are $N_i(0)= 5407$, $N_q(0) =0,N_{tr}(0)=0, N_r(0) = 0, N_d(0) =0$. (The reason for using zero initial conditions for $N_q(0),N_{tr}(0), N_r(0)$ and $N_d(0)$ is that the values of these populations would not be recorded during the first phase. However, using the values obtained from the first phase for $N_r$ and $N_d$ makes little difference. Note that $N_i(0)= 5407$ is smaller than the total number of infected cases. Again, using $N_i(0) = 5687$ does not alter the results.) The  parameters values used are those listed in Table \ref{Select-Set-FM}. Figure \ref{UK-Fullmodel}(a) shows plots of the calculated total infected population $N_t$ and the total infected cases in the UK ($\bullet$). Clearly, the fit is very good. The asymptotic saturation value of the total number of infections $N_t$ turns out to be $3.52 \times 10^5$. Also shown is the active infected $N_i$ labeled (ii). Further, the plot of the model predicted active infected population $N_i$ (ii) exhibits a turning point (peak) around May 15. Subsequent decrease in $N_i$ is seen to be slow. At this rate of slowing-down, the model predicts that a near saturation value of $3.52 \times 10^5$.

 Now consider  the estimation of the end time of the epidemic. A natural candidate for estimating this is the daily new infections. Considering  the close fit of the model predicted cumulative infected population $N_t$ with the UK data  (see Fig. \ref{UK-Fullmodel}a), the model calculated  daily new infections $\frac{dN_t}{dt}$ can be compared with the UK data for the daily new cases. This is  shown in Fig. \ref{UK-Fullmodel}(b). As can be seen the general profile of the model predicted daily new infected cases  matches  well with the published data. Recall that  the end time of the epidemic is defined as the time at which no new cases are reported. This definition is clearly   impractical for  forecasting  since the state of saturation is always reached asymptotically. For this reason, we use an arbitrarily small value, say 300 new cases as the end date of the epidemic. Then, the estimated  end time is late July.

To the best of our knowledge, we are not aware of any model that fits the COVID-19 data for any country as has been done here, particularly  over such long periods with the ability to forecast the future progression of the disease. However, there have been some efforts to fit data for the initial periods \cite{Zhao20,Chinta20,Tobias20,Tuite20,Sebastiani20,Gatto20,Guzze20}. In view of the good fit for the UK data, the model can be used to fit the COVID-19 data for other countries and also to forecast the progression of the disease.

\section{Summary, Discussion and Conclusions}

Recent literature has focused on abstracting the effect of various types of interventions through epidemiological models to make projections of how the disease progresses under different conditions. Recall that one limitation particularly applicable to the deterministic compartmental models is the difficulty in getting proper estimates of the parameters, particularly when the number of compartments is large. In this respect, simpler models with fewer compartments have an advantage. However, several factors may contribute to a single parameter. This is also  a model-dependent feature. Therefore the ability of such parameters to represent the mitigating efficacy of interventions appears limited (see below). Furthermore, the number of parameters in such models is not necessarily small, making numerical solutions often the only choice. Therefore, any method - whether mathematical or conceptual - which simplifies analysis and easy interpretation is welcome.

Motivated by this, we   hypothesize accessible population for transmission of the disease that can be the total or a small fraction of the total population depending on whether the transmission dynamics evolves in the absence or presence of interventions. Indeed, the effect of lock-down  is evident in all counties where the disease has been controlled or nearly eliminated. At the mathematical level, we introduce a decoupling scheme to aid mathematical analysis that also helps easy interpretation. The model equations have been devised in such a way that the susceptible and active infected populations form the main populations. The decoupling is affected by dropping all inward and outward transitions excepting the direct transitions ($ p_i \beta_i N_a(0)$ and $\alpha_s p_s N_a(0)$). Because, all outward transitions from $I$ are ignored under this decoupling, the active infected population $N_i$ takes the role of the cumulative infected population $ N_t$. The simplicity of the reduced logistic equation (Eq. \ref{C-active}) allows easy identification of the growth and inhibiting factors in terms of the dominant growth factors (direct inwards transitions or parameters). Surprisingly, this simple equation provides a good fit to the reported cumulative number of infections for the United Kingdom, as is clear in Fig. \ref{UK-Fit-logistic}(b).

The full model Eqs. (\ref{Sus}- \ref{Dead}) contain several parameters whose range has been estimated in a number of studies\cite{Zhao20,Chinta20,Tobias20,Tuite20,Gatto20,Dorigatt20,Anasta20,Wang20,Guzze20}. However, when it comes to explaining or capturing the growth characteristics for a specific country, optimized parameters suitable for the situation are required. Following \cite{Srikanth17,GA18,GA19}, we have determined the relative importance of the various transition rates (equivalently the associated parameters) subject to the constraint that the parameter values provide the best fit for the given data. In this work, we have made use of publicly available data on the total infected  and daily new infected cases  for the United Kingdom \cite{WHO}. 

Figure \ref{UK-Fullmodel}(a) shows the fit obtained for the period till March 23, 2020 (shown in the inset) and for the period beyond. Clearly, the fit is seen to be very good for both the period till the lock-down date and the period thereafter. Comparing Fig. \ref{UK-Fullmodel}(a) with Fig. \ref{UK-Fit-logistic}(b) for the reduced logistic map, we see that while the fits in both cases are comparable, the projections of the future progressions  are significantly different. The saturation value predicted by the full model (shown in Fig. \ref{UK-Fullmodel}(a) is close to $3.52 \times 10^5$, whereas that predicted by the reduced logistic equation in Fig. \ref{UK-Fit-logistic}(b) is $\sim 2.9 \times 10^5$. Conventionally, the end time of epidemic is defined as the day on which no new infections are reported. However, since the approach to the end point is generally slow, a better variable to predict the end point of the epidemic is by comparing  the model computed daily new case with the reported data on the daily new cases for the UK. Then, the end time of the epidemic predicted by the full model turns out be late July (see Fig. \ref{UK-Fullmodel}b). In contrast, the end time for the epidemic predicted by the reduced logistic model is late June. Clearly, the results obtained from the full model emphasize the limitations of the reduced model. A natural question is: what are the underlying causes?

The fact that the reduced logistic model provides a good fit also means that the major contributing factors for the growth of infection are included in Eq. (\ref{Infec}). To see this, consider Eqs. (\ref{Sus}-\ref{Dead}). The growth of $N_i(t)$ or more appropriately the daily new infected cases $\frac{dN_i}{dt}$ has two types of inward transitions, namely, direct and delayed. However, the dominant direct transition from $S$ to $I$ given by $ p_i \beta_i $ controls the growth rate of $N_i$. The other direct transition $\alpha_s p_s$ into $I$ also contributes to a lesser extent. Now, consider the delayed inward transitions to $I$ coming from $Q$ and $T$. These transitions are smaller in magnitude and contribute to sub-exponential growth of $N_i$ in time. More importantly, the turning point in $N_i$ or $\frac{dN_i}{dt}$ is due to a competition between the growth factors (all inward transitions) and the outward transitions (recovery and fatality terms). Further, since the time evolution beyond the turning point of $\frac{dN_i}{dt}$ is controlled by a balance between all inward and all outward transitions, the approach towards the state of no infections or the saturation value of $N_t$ is slow in our case. These features are clear from Fig. \ref{UK-Fullmodel}(a,b). Note that the fit, till June 1, is just two weeks beyond the turn point of $\frac{dN_i}{dt}$ and it has a long way to evolve to the end point of the epidemic. 

These arguments clarify two features of the data fit obtained using the reduced logistic equation. Because the total number of infected cases $N_t$ has the dominant growth contributions, the good fit is not surprising. On the other hand, growth dynamics beyond the turning point in the daily new infected cases (i.e., $\frac{dN_i}{dt}$) is controlled by a balance between growth factors (all inward transitions) and inhibiting factors (the rate of recovery and dead). However, these competing time scales are absent in the logistic equation. This clarifies why the projected saturation value of $N_t$ and the end time of the epidemic is not well captured.

A few comments above model are in order. As stated  in the introduction, the standard compartmental models ignore both age dependence and the heterogeneous spatial distribution of the population, and therefore  cannot abstract the aspects that depend fundamentally on these two features. For the same  reason, the populations in these models represent only the  mean response of each of these populations. From this point of view, the good fit obtained by the full model, to lesser extent by  the reduced logistic model, may come as a surprise. However, the success story of the mean field approaches has been established in physics literature \cite{Strogatz,Srikanth17,GA18,GA19} and the limitations of the approach has also been well established. To this extent, the success  of the present  model can be attributed to the way the model equations are structured and the underlying nonlinear dynamical methods used for analysis.

Here, we mention that the parameter values used for fitting have been obtained using an optimization procedure subject to the constraint that the optimized values should fit  the data  for the cumulative infected cases for the UK. Although, each parameter is varied within the range of values   estimated in  the literature (and open sources), since the optimization has been carried out subject to only one constraint, the  values may not be unique. From this point of view, more number of constraints  such as the data for  the active infected, recovered, quarantines etc., would be helpful in removing the non-uniqueness of the optimized values, at least partially.

As stated in the text, the contact transmission rate parameter $\beta_i$ is one of the crucial parameter in the model because this parameter largely controls the time development of the disease. In our approach, this  parameter has been estimated by using the local initial slope in the log-linear plot of the UK data. The slope itself, however, depends on whether the disease development occurs in the absence or presence of interventions, which in turn depends on the accessible population $N_a(0)$. Since the model growth rate depends on $p_i\beta_i N_a(0)$, the value of $\beta_i$ depends  inversely on $N_a(0)$  corresponding to the  absence or presence of interventions. However, the value reported by Tang {\it et al.} \cite{Tang20a} is two orders smaller than  that estimated in our paper. However, in both cases, the value of $\beta_i$ is inversely proportional to the relevant population used for modeling. In this context, we mention that there is only independent estimate both under free evolution \cite{Tang20}. Such an independent estimates are  desirable. 

One point that needs some discussion is about the values of the recovery $\gamma_r$ and the death $\kappa_d$ rates  obtained from the optimization procedure used that fit the UK data very well (see Table \ref{Select-Set-FM}). These rates are inversely related to the time
duration between detection of illness till recovery and death respectively. While these values  are within the published range of values \cite{Dorigatt20,Anasta20,Wang20}, they appear to be on the lower side of the mean.  However,  in principle, the parameter values depend on the structure of model equations. In our model, the recovery and fatality are outward transitions from a single compartment, namely, $I$ to $R$ and $D$.  This feature is clearly because our model equations were structured to have only two  core populations. However, the recoveries can occur  if other kinds of compartments (populations) are included \cite{Tang20}.  Thus, the best fit for lower values may be the result of simplicity of the model.

In conclusion, the simple compartmental model not only provides a good fit to the United Kingdom coronavirus data but also makes concrete long term predictions for the future. We believe that these results have been made possible due to the reductive approach adopted here.

%\appendix*
\appendix*
\section{}

Recall the equation governing the cumulative infected population $N_s(t)$ from Eqs. (\ref{Acc}-\ref{Cinfec}) is given by 
\begin{eqnarray} 
%\nonumber
\label{C-active1}
{\dot N}_t &=& c + b{N}_t- a {N}_t^2,\\
\label{A1}
a &= & p_i \beta_i, \\
\label{B1}
b &= & p_i \beta_i N_s(0) -\alpha_s p_s,\\ 
\label{C1}
 c &= & \alpha_s p_s N_s(0).
\end{eqnarray}
Equation (\ref{C-active1}) has the well known form of the logistic equation extensively studied in the context of population dynamics. However, the parameters $a,b,$ and $c$ have a well defined interpretation.

 Now consider the solution of Eq. (\ref{C-active1}). Let $\alpha_{1,2}=\frac{b \pm \sqrt{b^2+4ac}}{2a}$ be the roots of the quadratic equation. Then, in terms of $a, b$ and $c$, the two roots can be written as $\alpha_1 \sim \frac{b}{a} = N_s(0)$ and $\alpha_2 \sim - ac/b < 0$, which is small compared to $b$. Then the solution is given by 
 
\begin{equation}
{ N}_t = \frac{A\alpha_1 e^{ a(\alpha_1-\alpha_2)t }-\alpha_2 }{Ae^{a(\alpha_1-\alpha_2)t }-1} =\frac{A\alpha_1 e^{ bt } -\alpha_2}{Ae^{bt} -1}.
\label{Full-solu1} 
\end{equation}

 The constant $A $ is given by 
\begin{equation}
 A=\frac{{ N}_t(0)-\alpha_2}{{N}_t(0)-\alpha_1}. 
\label{AA}
\end{equation}
Then, we have 
\begin{equation}
{ N}_t = \frac{\big(\frac{b}{a} {N}_it0) + \frac{c}{a} \big) e ^{\, b t} + \frac{c}{a} + \frac{ac}{b}}{\big( { N}_t(0) + \frac{ac}{b} \big) e ^{\, b t} - { N}_t(0) + \frac{b}{a} }.
\label{Full-solu2} 
\end{equation}
For short times, ${ N}_t$ tends to $ ({N}_t(0) + \frac{c}{b} \big) e ^{\, b t}$ (since the denominator is dominated by $b/a = N_s(0)$), consistent with Eq. (\ref{Linear-Ni}), the short time solution. For long times however, ${N}_t$ tends to $b/a = N_s(0)$, the total population. 
\iffalse
\begin{acknowledgements}
G.A. acknowledges Indian National Science Academy for a Honorary Emeritus Scientist position. We thank Professor S. A. Shivashankar for useful discussions.
%If you'd like to thank anyone, place your comments here
%and remove the percent signs.
\end{acknowledgements}

% Authors must disclose all relationships or interests that 
% could have direct or potential influence or impart bias on 
% the work: 
%
 \section*{Conflict of interest}
%
 The authors declare that they have no conflict of interest.

% BibTeX users please use one of
%\bibliographystyle{spbasic}      % basic style, author-year citations
%\bibliographystyle{spmpsci}      % mathematics and physical sciences
%\bibliographystyle{spphys}       % APS-like style for physics
%\bibliography{}   % name your BibTeX data base

% Non-BibTeX users please use
\fi

\end{document}